\documentclass[fleqn,usenatbib]{mnras}
\usepackage{newtxtext,newtxmath}
\usepackage[T1]{fontenc}

\DeclareRobustCommand{\VAN}[3]{#2}
\let\VANthebibliography\thebibliography
\def\thebibliography{\DeclareRobustCommand{\VAN}[3]{##3}\VANthebibliography}

\usepackage{graphicx}	
\usepackage{amsmath}	
\usepackage[nolist]{acronym}
\usepackage[noabbrev]{cleveref}
\usepackage[range-units = single]{siunitx}
\usepackage{booktabs}

\newcommand\m{\mathrm}

\title[Parameter constraints for AMPs]{Parameter constraints for accreting millisecond pulsars with synthetic NICER data}

\author[Dorsman et al.]{Bas Dorsman,$^{1}$\thanks{E-mail: b.dorsman@uva.nl} 
Tuomo~Salmi,$^{1,2}$
Anna~L.~Watts,$^{1}$
Mason Ng,$^{3,4,5}$
Satish Kamath,$^{6}$
Anna Bobrikova,$^{7}$\newauthor
Juri Poutanen,$^{7,8}$
Vladislav Loktev,$^{7,2}$
Yves Kini,$^{1}$
Devarshi Choudhury,$^{1}$
Serena Vinciguerra,$^{1}$\newauthor
Slavko Bogdanov$^{9}$
and Deepto Chakrabarty$^{3}$
\\
$^{1}$Anton Pannekoek Institute for Astronomy, University of Amsterdam, Science Park 904, 1098XH Amsterdam, the Netherlands\\
$^{2}$Department of Physics, P.O. Box 64, FI-00014 University of Helsinki, Finland\\
$^{3}$MIT Kavli Institute for Astrophysics and Space Research, Massachusetts Institute of Technology, 77 Massachusetts Avenue, Cambridge, MA 02139, USA\\
$^{4}$Department of Physics, McGill University, 3600 rue University, Montr\'{e}al, QC H3A 2T8, Canada\\
$^{5}$Trottier Space Institute, McGill University, 3550 rue University, Montr\'{e}al, QC H3A 2A7, Canada\\
$^{6}$SURF, Science Park 140, 1098 XG Amsterdam, the Netherlands\\
$^{7}$Department of Physics and Astronomy, FI-20014 University of Turku, Finland\\
$^{8}$Space Research Institute, Russian Academy of Sciences, Profsoyuznaya 84/32, Moscow 117997, Russia \\
$^{9}$Columbia Astrophysics Laboratory, Columbia University, 550 West 120th Street, New York, NY 10027, USA
}

\date{Accepted XXX. Received YYY; in original form ZZZ}

\pubyear{\the\year{}}

\def\largespot{Scenario A}
\def\smallspot{Scenario B}

\def\scenario{scenario}

\def\scenarios{scenarios}

\acrodefplural{EoS}{equations of state}
\acrodefplural{LMXB}{low-mass X-ray binaries}

\begin{acronym}
\acro{1D}{one dimensional}
\acro{2D}{two dimensional}
\acro{3D}{three dimensional}
\acro{4D}{four dimensional}
\acro{5D}{five dimensional}
\acro{AMP}{accreting millisecond pulsar}
\acro{EoS}{equation of state}
\acro{GR}{general relativistic}
\acro{MHD}{magneto-hydrodynamic}
\acro{GW}{gravitational wave}
\acro{HPC}{high-performance computing}
\acro{J1808}{SAX J1808.4$-$3658}
\acro{LMXB}{low mass X-ray binary}
\acro{MP}{millisecond X-ray pulsar}
\acro{NICER}{Neutron Star Interior Composition Explorer}
\acro{NS}{neutron star}
\acro{PPM}{pulse-profile modelling}
\acro{RMP}{rotation powered millisecond pulsar}
\acro{ST}{single temperature}
\acro{TBO}{thermonuclear burst oscillation}
\acro{QPO}{quasi-periodic oscillation}
\acro{X-PSI}{X-ray Pulse Simulation and Inference}
\end{acronym}

\crefformat{equation}{equation~#2#1#3}

\begin{document}
\label{firstpage}
\pagerange{\pageref{firstpage}--\pageref{lastpage}}
\maketitle

\begin{abstract}
Pulse profile modelling (PPM) is a technique for inferring mass, radius and hotspot properties of millisecond pulsars. PPM is now regularly used for analysis of rotation-powered millisecond pulsars (RMPs) with data from the Neutron Star Interior Composition ExploreR (NICER). Extending PPM to accreting millisecond pulsars (AMPs) is attractive, because they are a different source class featuring bright X-ray radiation from hotspots powered by accretion. In this paper, we present a modification of one of the PPM codes, X-PSI, so that it can be used for AMPs. In particular, we implement a model of an accretion disc and atmosphere model appropriate for the hotspots of AMPs, and improve the overall computational efficiency. We then test parameter recovery with simulated NICER data in two scenarios with reasonable parameters for AMPs. We find in the first scenario, where the hotspot is large, that we are able to tightly and accurately constrain all parameters including mass and radius. In the second scenario, which is a high inclination system with a smaller hotspot, we find slightly widened posteriors, degeneracy between a subset of model parameters, and a slight bias in the inferred mass. This analysis of synthetic data lays the ground work for future analysis of AMPs with NICER data. Such an analysis could be complemented by future (joint) analysis of polarization data from the {\it Imaging X-ray Polarimetry Explorer} ({\it IXPE}).
\end{abstract}


\section{Introduction}
\subsection{Theory and context}
\Acp{NS} are the densest astrophysical objects in which matter still resists the gravitational pressure to collapse into black holes. Because of this, their cold dense cores provide unparalleled physics laboratories. Which particles are present there, and their interactions, are not yet well understood \citep[see e.g.][for reviews]{Lattimer2016, Burgio2021}. 

A macroscopic descriptor of the interior of \acp{NS} is the \ac{EoS}: the relationship between density and pressure throughout the star. Constraining the \ac{EoS} can be done via nuclear theory \citep[e.g.][and references therein]{Leonhardt2020} and laboratory experiments on Earth \citep{Adhikari2021, Adhikari2022}. Besides these, the \ac{EoS} is also a major topic of research in modern astrophysics \citep{Chatziioannou24}. It is being measured via radio timing of high mass pulsars \citep[e.g.][]{Demorest2010, Antoniadis2013, Cromartie2020, Fonseca2021}, 
via spectral evolution of X-ray burst emission \citep[e.g.][]{Suleimanov2011,Nattila2016,Nattila2017}, 
via \acp{GW} that probe the tidal deformability of binary \ac{NS} mergers \citep{Abbott2019, Abbott2020}, and via \ac{PPM} of rapidly rotating \acp{NS} \citep{Poutanen2003,Watts2019}. The last of these will be the focus of this work. 

\ac{PPM} is the technique which simulates energy and phase resolved pulse profiles, which are produced by the rotation of a \ac{NS} with radiating surface patterns. Then, typically via sampling the parameter space and comparing expected pulse profiles against actual data, one can estimate posterior probability distributions of model parameters. In X-ray pulsars, inference of the compactness ($M/R$) of the star is enabled because it is encoded in the light bending of beamed radiation emergent from hotspots \citep{Pechenick1983, Riffert1988}. For rapidly rotating (i.e. millisecond) pulsars, it is possible to harness the special and general relativistic effects to measure both mass and radius individually, while also taking into account oblateness \citep[][]{Miller1998, Poutanen2003, Poutanen2006b, Morsink2007, Baubock2013, AlGendy2014, Psaltis2014}. The \ac{EoS} is then informed by these mass and radius measurements because they are linked via the stellar structure equations. Besides $M$ and $R$, surface features -- which are connected to the magnetic field structure\footnote{They are at least for rotation and accretion powered millisecond pulsations but the connection is less clear for thermonuclear burst oscillations \citep[see e.g.][]{Watts2012}.} -- are also of interest \citep[e.g.][]{Kalapotharakos2021, Das2022, Petri2023}.

Currently, \ac{PPM} is being used to constrain the masses and radii of \acp{RMP}, which are the primary targets of the \ac{NICER}. These include PSR J0030+0451 \citep{Riley2019, Miller2019, Salmi2023,Vinciguerra2024} and PSR J0740+6620 \citep{Riley2021, Miller2021, Salmi2022, Salmi2023,Dittmann2024, Salmi2024}, and more recently also PSR J0437--4715 \citep{Choudhury2024a}. The modelling of pulse profiles has been well tested \citep[e.g.][]{Bogdanov2019b, Choudhury2024b} and the Bayesian parameter inference has been cross-verified \citep{Bogdanov2021} as well as tested independently \citep{Afle2023}.

\ac{PPM} is also starting to be applied to two types of pulsations seen in \acp{NS} in \acp{LMXB}. The first are thermonuclear burst oscillations \citep{Kini2023,Kini2024a,Kini2024b}. In this case the accumulation of accreted matter from the binary companion on the \ac{NS} surface leads to bright thermonuclear bursts during which pulsations are sometimes present at or very close to the rotational period of the \ac{NS}  \citep[for reviews, see e.g.][]{Watts2012, Galloway2021}. The second are the pulsations from the persistent hotspots on the \ac{AMP} surface and these are the focus of this work.

\acp{AMP} \citep[for reviews, see][]{Poutanen2006a, DiSalvo2022}, discovered in 1998 by \cite{Wijnands1998}, provided confirmation of the recycling scenario. This scenario explains the fast rotation of millisecond pulsars via angular momentum transfer during accretion from a binary companion \citep[e.g.][]{Alpar1982, Bhattacharya1991}. While \acp{RMP} are the leftover result after this process has taken place, \acp{AMP} are a younger population that are being caught in the act of spinning-up. This makes \acp{AMP} an interesting population to study from the perspective of X-ray binary evolution.
Nevertheless, in \aclu{J1808} (hereafter J1808) it has been observed that when following the spin evolution across several outbursts, the neutron star appears to be spinning down (e.g. \citealt{Hartman2008, Illiano2023}). This seeming contradiction motivates the need to understand spin-up and spin-down mechanisms and their interplay more comprehensively across pulsars' lifecycles.

Being powered by accretion, \acp{AMP} bear witness to a very different physical environment compared to \acp{RMP}. In \acp{AMP}, the pulsations are produced by surface hotspots which are heated by the bombardment of accreted material. This material is typically being funneled along the magnetic field lines onto the magnetic poles of the star. As photons escape from the hotspot, some undergo inverse Compton scattering due to hot accreted electrons. The phase-averaged X-ray spectrum of an \ac{AMP} is typically composed of one or two blackbody-like components (usually associated with the \ac{NS} surface and an accretion disc) and a Comptonized component (from the hotspot, \citealt{DiSalvo2022}). In some cases broadened reflection lines from the inner region of the disc are detected, which may help constrain the inner disc radius (\citealt{Cackett2009, Papitto2010, Wilkinson2011, DiSalvo2019} or for a recent review \citealt{Ludlam2024}). The X-ray pulses have been observed to vary over the course of an outburst, bearing witness to the complex interplay between the magnetic field, accretion rate and the disc throughout an outburst \citep{Hartman2008,Poutanen2009, Ibragimov2009, Kajava2011, Bult2020}. In addition, it has recently been observed that \acp{AMP} can produce polarized radiation through inverse Compton scattering \citep{Papitto2025}. The polarization degree and angle highly depend on the scattering angle and electron temperature \citep{Nagirner1994} and carry information about the surface geometry of hotspots \citep[e.g.][]{Viironen2004}.

\subsection{Previous studies of pulse profiles of AMPs} 

Despite the complexity, a considerable amount of work has been done to study the pulse profiles of these objects. \cite{Poutanen2003} used {\it RXTE} data to fit the pulses of \ac{J1808} using two broad energy bands (3--4 and 12--18~keV) using the chi-square method. The model for the star included Comptonized and blackbody radiation from the hotspot, Doppler boosting, relativistic aberration and gravitational light bending in the Schwarzschild space–time. The accretion disc is also included in their model in the form of spectral contributions from Compton reflection and Fe lines (6--7 keV). From the pulse profiles they infer a restriction on the allowed mass range of 1.2 to 1.6\,$M_\odot$ and infer respective radii of 6.5 to 11~km.

\cite{Leahy2008} also inferred the mass and radius of \ac{J1808} with a modified space-time model that includes stellar oblateness and light travel time delay across the star. Other similar analyses include \citet[][XTE J1751$-$305]{Gierlinski2005}, \citet[][XTE J1814$-$338]{Leahy2009}, and \citet[][for multiple epochs of \ac{J1808}]{Morsink2011}.

More recently, \cite{Salmi2018} put Bayesian parameter constraints on \ac{J1808} with energy and phase resolved {\it RXTE} data. They were able to get accurate and precise mass-radius constraints using synthetic data, with improved precision if star and spot inclination were known. However, with real data and broad priors on mass and radius they found smaller masses and radii than is expected for modern \acp{EoS}. Still, by using a fixed mass grid they found reasonable radius constraints. They included a model for the Compton reflection off the accretion disc, the Fe lines, \texttt{xilconv} \citep{Garcia2013}, and an empirical model for the Comptonization spectrum, \texttt{SIMPL} \citep{Steiner2009}. They parameterized the angular dependence of the Comptonized radiation so that the intensity has a linear dependence on $\cos\alpha$, where $\alpha$ is the emission angle. This may be a source of error, especially at higher energy.

Earlier, \cite{Viironen2004} had set the stage for the study of X-ray polarization data from accretion powered pulsations and \acp{TBO}. This was built upon by \cite{Salmi2021} who performed \ac{PPM} using simulated polarization data from the {\it Imaging X-ray Polarimetry Explorer} ({\it IXPE}), using the Thomson scattering approximation in an optically thin \ac{NS} atmosphere. 

Hotspot shape, size, and position have also been investigated through \ac{3D} \ac{MHD} simulations for example by \cite{Romanova2004} and \cite{Kulkarni2013}, who found dependencies on parameters such as accretion rate and disk inclination. Additionally, the investigation of light scattering in the accretion funnel by \cite{Ahlberg2024} has identified the significance of scattering on the pulse profiles, with its impact varying based on the accretion rate.

\begin{figure*}
\centering
\includegraphics[width=0.9\textwidth]{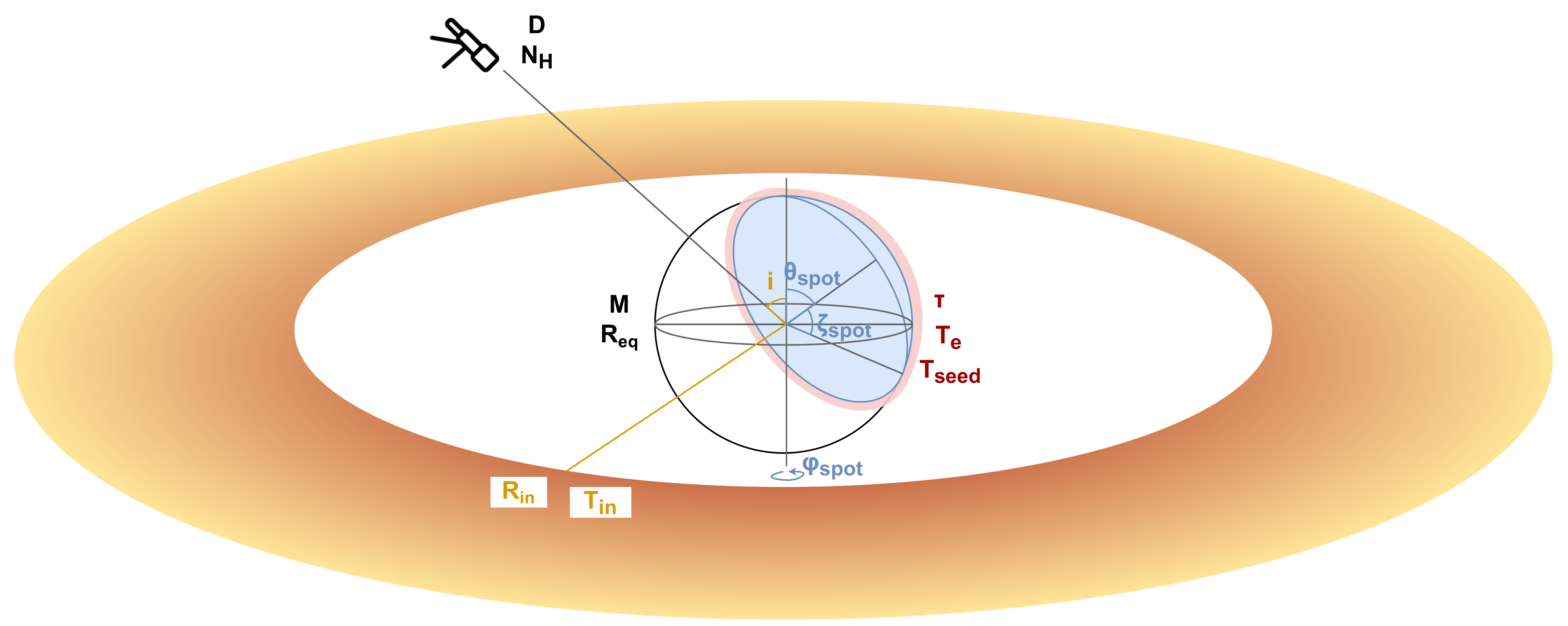}
\caption{Diagram of the \ac{AMP} model. In blue, the hotspot co-latitude $\theta$, angular radius $\zeta$ and phase $\phi$ are given. In black, star mass $M$, equatorial radius $R_{\rm eq}$,  distance $D$ and neutral hydrogen column density $N_{\rm H}$ are given. In yellow the observer inclination $i$, inner disc radius $R_{\rm in}$ and inner disc temperature $T_{\rm in}$ are given. Finally in red the atmosphere parameters are shown: optical depth $\tau$, electron temperature $T_{\rm e}$ and seed photon temperature $T_{\rm seed}$.}
    \label{fig:diagram}
\end{figure*}

\subsection{This work}
This study investigates the prospects for future parameter inference of \acp{AMP} with \ac{NICER} data. \ac{NICER} is an excellent timing instrument that has already observed and discovered \acp{AMP} \citep[e.g.][]{Bult2020, Ng2021, Bult2021, Ng2024}. It is sensitive in an energy band (0.2 to 12 keV) complementary to for example {\it RXTE} (2 to 60 keV for the Proportional Counter Array) which was previously used to study \acp{AMP} \citep[e.g.][]{Salmi2018}. Because of this, \ac{NICER} probes the expected lower energy contribution from the accretion disc \citep{Patruno2009a} in more detail. \ac{NICER} is also conducting joint observations of \acp{AMP} with the X-ray polarimetric mission {\it IXPE} \citep{Papitto2025}. While in this work we assess first what might be achievable with \ac{NICER} alone, in a complementary analysis, \cite{Salmi2025}, infer parameters from pulse profiles of synthetic {\it IXPE} data in various scenarios, including the two scenarios that we will consider in this paper. We note that a future study may analyse both \ac{NICER} and {\it IXPE} data jointly, possibly leading to enhanced parameter constraints.

We examine the precision and accuracy of parameter recovery attainable with synthetic \ac{NICER} data,\footnote{With precision we mean whether the width of the posterior distribution is small, while with accuracy we mean whether the true input parameter is contained within that posterior, e.g. within some appropriate credible interval.} and if there are biases that may be inherent to the analysis. We consider two reasonable \ac{AMP} \scenarios{} with different hotspot and inclination configurations.

To do this we introduce two new model components to the established \ac{X-PSI} software (one of the \ac{PPM} packages that has been used in analysis of NICER \ac{RMP} data, see \citealt{Riley2023} and  \Cref{sec:method}). The first is a new emission model appropriate for an accreting hotspot (\Cref{sec:atmosphere}) developed by \cite{Bobrikova2023}. Compared to the model for the spectrum and angular dependence employed by \cite{Salmi2018}, emission is now based on thermal Comptonization of seed photons by hot electrons in an isothermal plane-parallel slab geometry at the \ac{NS} surface. The second is a blackbody component from the accretion disc (\Cref{sec:disc}). This was not included in previous analyses with {\it RXTE} data because it is expected below the lower limit of the {\it RXTE} band at $\sim$2 keV. However, it should be well within the \ac{NICER} band.

The introduction of new model components increases the computational expense of the likelihood evaluation in \ac{X-PSI}. Because this computational expense is a bottleneck when doing Bayesian sampling to estimate posterior distributions, we make efforts to reduce the computational costs. Most progress is made by optimizing the interpolation associated with the atmosphere model.

This paper has been structured as follows. \Cref{sec:model} describes the models that produce the pulse profiles in this work. \Cref{sec:optimisation} describes the newly introduced computational optimization in the implementation of the model in \ac{X-PSI}. \Cref{sec:method} describes the methodology applied, i.e. Bayesian parameter inference, and introduces the two synthetic datasets (which we call `scenarios') to which this analysis is applied. \Cref{sec:results} presents the results of the parameter inference, and these are discussed further and put into context in \Cref{sec:discussion}. Finally, \Cref{sec:conclusion} summarizes and concludes the paper.

\section{Model}\label{sec:model}
We use \ac{X-PSI},\footnote{We use \ac{X-PSI} v2.2.5. The \href{https://github.com/xpsi-group/xpsi}{Github repository} provides a working \href{https://github.com/xpsi-group/xpsi/blob/disk/examples/examples_modeling_tutorial/TestRun_AMXP.py}{example} of the model of an \ac{AMP} with accretion disc.} an open-source software for forward modelling of \ac{NS} X-ray pulses and inference of model parameters \citep{Riley2023}. \ac{X-PSI} was designed with the specific focus of modelling \acp{RMP} and has been extensively used for this purpose with \ac{NICER} data \citep[e.g.][]{Riley2019, Vinciguerra2024, Salmi2024, Choudhury2024a} with independent replication by \citet{Afle2023}. This section describes the model with a focus on the newly added \ac{AMP} components. \Cref{fig:diagram} displays a diagram of the model which visualizes the parameters.

\subsection{Spacetime and ray-tracing}
The model of the \ac{NS} uses the Oblate Schwarzschild approximation \citep{Morsink2007, AlGendy2014} which adds oblateness of the star to the previously used Schwarzschild and Doppler approximation \citep{Miller1998, Poutanen2003}. Ideally, the rotating \ac{NS} would be embedded in a spacetime that was solved according to numerical methods, but this would be prohibitively expensive (especially when sampling) and the Oblate Schwarzschild approximation is accurate enough for the spin rates of all known accreting \acp{NS} \citep{Silva2021}.  

To give a rough overview of the ray tracing through the spacetime, the gravitational light bending is computed first assuming a Schwarzschild metric for a grid of angles. Then through interpolation one can compute the surface angles of the rays from the oblate star that eventually hit a given observer. Transforming between observer and star frame involves correcting the emission angle for special relativistic aberration and correcting the energy and intensity of the radiation for special relativistic and general relativistic effects. Flux received by the observer is a sum over contributions from the cells into which the surface hotspot is discretised. Time lags between light travel paths are also taken into account when computing phase dependent flux. This methodology to simulate the emission is described in more detail in \cite[e.g.][]{Pechenick1983, Miller1998, Poutanen2003, Bogdanov2019b}.

\subsection{Atmosphere}\label{sec:atmosphere}

The intensity of photons escaping the surface is described by a beaming function, which is a function of the cosine of the zenith angle $\mu$, photon energy $E$ and parameters governing the \ac{NS} atmosphere. In the case of \acp{AMP}, due to the material being accreted onto the \ac{NS}, the surface is heated and some of the radiation emitted from the surface is Comptonized by   electrons heated by the accretion process. We adopt the emission profile computed by \cite{Bobrikova2023}, who computed radiative transfer through a plane parallel slab geometry of hot electrons. There are three atmosphere parameters: seed photon temperature $T_{\rm seed}$, electron temperature $T_{\rm e}$ and Thomson optical thickness $\tau$. Rather than solving the radiative transfer while sampling, we interpolate from pre-computed tables, which saves on computational cost. This interpolation is an unchanged methodology compared to previous \acp{RMP} parameter inference efforts with X-PSI \citep[e.g.][]{Riley2019} with the NSX atmosphere \citep{Ho2001, Ho2009}, which is computationally cheaper given that it only has two parameters. 

It has been standard practice in \ac{X-PSI} analyses so far to assume that hotspot parameters stay constant along the azimuthal direction, which allows for a computational short-cut to be taken in signal integration. In the case of \acp{RMP}, the effective temperature $T_{\rm eff}$ is homogeneous across the spot, while the surface gravity $g$ is only constant along azimuth, and varies slightly with co-latitude due to the oblateness of the star. In the case of \acp{AMP}, we assume all the atmosphere parameters to be constant across the hotspot in both directions, which enabled us to apply a further optimization to the computational efficiency of the signal integration. This is detailed in \Cref{sec:optimisation}.

\subsection{Hotspot geometry}\label{sec:geometry}

The hotspot geometry of \acp{AMP} is determined by the cross-sectional shape of the accretion flow bombarding the \ac{NS} surface, which is not well known. In this work, we use the most simple hotspot geometry, which is a single spherical cap with atmosphere parameters homogeneous across the hotspot. Outside the hotspot we assume zero flux. This is a very simplified picture, but it is a well-established starting point to parameterize the hotspot, previously used for example by \citet{Poutanen2003}, \citet{Salmi2018} and \citet{Bobrikova2023}. We discuss the caveats with this simplified picture, and potential improvements, in \Cref{sec:caveats}.

This circular shape is parameterized with just three parameters: hotspot co-latitude $\theta$, angular radius $\zeta$, and phase $\phi$. We assume for simplicity that the phase of the hotspot is constant. While this can be the case (it was for example during a few days after the peak of the accretion outburst of \ac{J1808} discussed by \citealt{Bult2020}), it may not always be valid -- phase drift in \acp{AMP} has often been observed and is most probably related to either accretion torque or movement of the hotspot on surface \citep{Hartman2008, Patruno2010, Kulkarni2013}.

\subsection{Accretion disc model}\label{sec:disc}
Due to the angular momentum carried by the matter being accreted from the binary companion, an accretion disc forms. For \acp{AMP}, the disc is truncated at the magnetospheric radius $R_{\rm m}$ within which charged matter follows the magnetic field lines and accretes onto hotspots on the \ac{NS} surface. Observations of \ac{J1808} (for example) indicate the presence of an accretion disc through Fe lines at 6--7 keV \citep{Papitto2009, Cackett2009}. Below $\sim$2 keV, the pulse fraction reduces due to the presence of an unpulsed component which is likely associated with the disc \citep{Patruno2009a}. Because \ac{NICER} is sensitive in this waveband, it is necessary to include it for \ac{PPM} with \ac{NICER} data.

The disc spectral model we use in this work is \texttt{diskbb} \citep{Mitsuda1984, Makishima1986}. The flux per energy 
is given by
\begin{equation}
\label{eq:disc}
f_\m{disc}(E) = \frac{8\pi R_\m{in}^2 \cos{i}}{3 D^2 T_\m{in}}\int_{T_\m{out}}^{T_\m{in}}\Big(\frac{T}{T_\m{in}}\Big)^{-11/3}B(E,T) dT.
\end{equation}
This is an integral that adds blackbody contributions from the disc rings, from the cold outer edge ($T_\m{out}$) to the hot inner edge ($r_\m{in}, T_\m{in}$). $B(E,T)$ is the Planckian distribution, given by
\begin{equation}
    B(E,T) = \frac{E^3}{c^2h^2}\left[\exp{\left(\frac{E}{k_\m{B}T}\right)-1}\right]^{-1}. 
\end{equation}

\subsection{ISM}

The attenuation of X-ray flux due to the interstellar medium is energy dependent. Here, as usual for \ac{NICER} analyses, the attenuation factor is solely parameterized by the column density $N_\mathrm{H}$ of neutral hydrogen. We use relative abundances for the interstellar medium from \cite{Wilms2000}. Their model \texttt{tbnew}\footnote{\url{https://pulsar.sternwarte.uni-erlangen.de/wilms/research/tbabs/}} provides pre-computed attenuation factors as a function of photon energy. This is described in more detail in Section 2.4 of \cite{Bogdanov2021}.

\section{Computational optimisation}\label{sec:optimisation}

\begin{figure}
\centering
\includegraphics[width=0.9\linewidth]{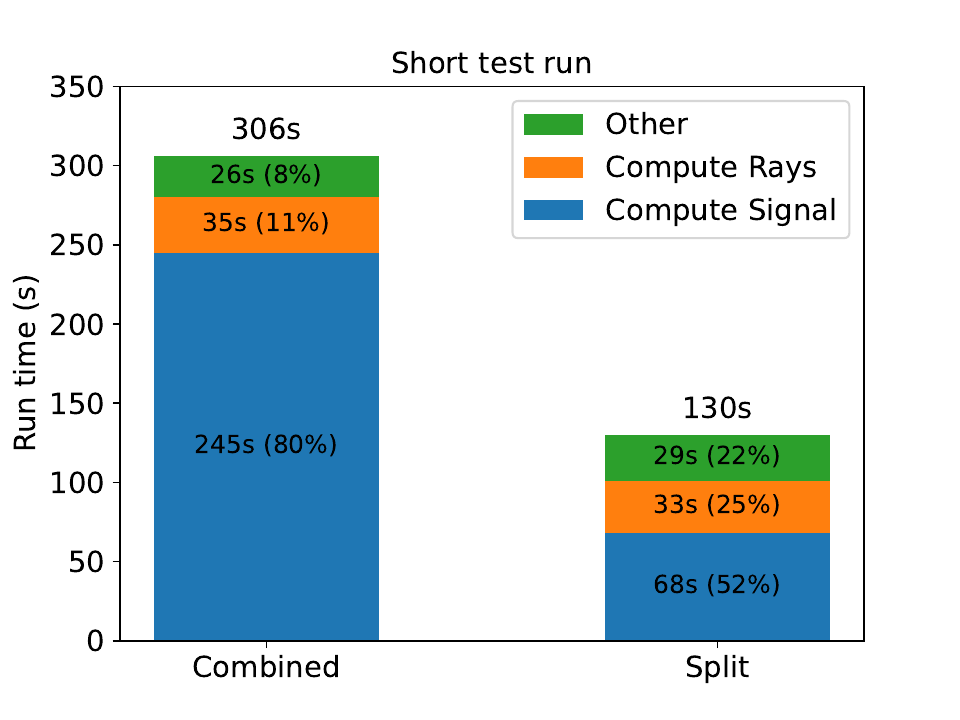}
\caption{The results of a short run test on a single core, where the sampling cuts off after 100 iterations. The left bar shows the results for the combined interpolation, and the right bar those for the split interpolation. Computation times are given in seconds and in percentage of the total computation time for each interpolation method. All model features and settings are identical between the left and right plot, except for the interpolation method. We recommend the reader to perform their own timing measurements when examining other models, because model features such as resolution settings and number of hotspots will affect these timing results.}
    \label{fig:run_test}
\end{figure}

With the complexity increase in modelling \acp{AMP} compared to \acp{RMP}, several efforts have been made to optimise the code.  
To start, we obtained a performance improvement of a few per cent by switching from Intel compilers to Free and Open Source Software (FOSS) compilers on the AMD hardware of the \ac{HPC} cluster Snellius.\footnote{The latest \ac{X-PSI} documentation includes installation instruction for HPC systems where FOSS based modules are used, for example in the case of the Dutch national supercomputer Snellius: \url{https://servicedesk.surf.nl/wiki/display/WIKI/Snellius}.}
We then profiled the code\footnote{We used \texttt{py-spy}, i.e. a sampling profiler for Python, available at \url{https://github.com/benfred/py-spy} and \texttt{speedscope}, i.e. a viewer for the resulting flamegraphs, available at \url{https://github.com/jlfwong/speedscope}.} and found that the largest contributor to computation time was the signal computation. \Cref{fig:run_test} summarizes the results of this test in the left bar. 

We note here that these are single thread tests, whereas many threads are typically used in a production run. From preliminary testing we found that with many threads a computational efficiency loss of a few percent can be expected, because the main process must wait for the slowest likelihood evaluation before it can continue and distribute new tasks. This is achieved via \textsc{MPI} barrier synchronization (see also Section 3.4.3 of \citealt{Riley2019thesis}).

The signal computation consists mostly of the interpolation required to compute the specific intensity emergent from the hotspot, and takes up a majority of the computational time. For atmospheres, the interpolation method is cubic multivariate Lagrange polynomial interpolation \citep[see e.g.][]{Sauer1995}. The code is implemented in an efficient manner in \textsc{Cython}, for example making use of cached values so that some computation steps can be skipped if a parameter is close enough to a previous value in any one of the dimensions. With our resolution settings (see \Cref{tab:xpsi_settings}), typically around 60,000 interpolations are performed for one model evaluation. This number corresponds roughly to the square root of number of cells the surface is discretised into (\texttt{sqrt\_num\_cells}) times the number of energies at which photon flux is calculated (\texttt{num\_energies}) times the number of rotational steps cell mesh motion is discretised into (\texttt{num\_leaves}).

The Compton slab atmosphere (see \Cref{sec:atmosphere}) model has five parameters that determine the emission intensity: $E$ (photon energy), $\mu$ (cosine emission angle), $T_{\rm seed}$ (seed photon temperature), $T_\m{e}$ (electron temperature) and $\tau$ (optical thickness), compared to NSX's usual four: $E$, $\mu$, $g$ (surface gravity) and $T_{\rm{eff}}$ (effective surface temperature). Interpolating in five dimensions is more expensive, but the additional expense has been mitigated by a redesign of the interpolation code. Now, we split the interpolation into two steps. First, because atmosphere parameters $T_\m{seed}$, $T_\m{e}$ and $\tau$ are constant across the hotspot, we use those constant values and perform \ac{5D} interpolations covering all tabulated values of $E$ (in our case 150 values) and $\mu$ (9 values). In doing so we create a secondary data table with $150 \times 9 = 1350$ intensity values on the fly. Second, to compute the actual intensities we interpolate $E$ and $\mu$ from the secondary table. While the creation of the secondary table is extra work (1350 \ac{5D} interpolations) which will have to be redone for each sampled hotspot, it effectively converts $\sim$60,000 \ac{5D} interpolations into the same amount of \ac{2D} interpolations, which are sufficiently cheaper such that we save time overall. The error introduced by this method is negligible: the intensities are identical up to $\sim$16 digits, the same level as double precision floats. After this optimization, we find through \texttt{py-spy} profiling the total run time has been reduced by more than half, as can be seen in \Cref{fig:run_test}. For additional context, \Cref{tab:test_signal} compares the time to compute the signal in both the old and new interpolation methods also against the time to compute the signal with a four parameter atmosphere used for \acp{RMP}. We see there that the new method roughly compensates for the additional atmosphere parameter. Note that it is not possible to optimize the \ac{RMP} atmosphere without losing some accuracy, because one parameter, the surface gravity, is not constant across colatitude.
 
\begin{table}
    \centering
    \begin{tabular}{cc} \hline 
        \hline
        \textbf{Setting}&
         \textbf{Value}\\ 
         \hline
    \texttt{sqrt\_num\_cells}& 50 \\ 
         \texttt{num\_leaves}& 30 \\ 
         \texttt{num\_energies}& 40\\ 
         \texttt{num\_rays}& 512\\ \hline
    \end{tabular}
    \caption{X-PSI settings used in this work. Their meanings are briefly described in \Cref{sec:optimisation}.}
    \label{tab:xpsi_settings}
\end{table}

\begin{table}
    \centering
    \begin{tabular}{ccc}         \hline \hline 
         \textbf{Parameters}&
         \textbf{Method}&
         \textbf{Compute Signal (s)}\\
         \hline 
         4 &combined &0.063\\ 
         5 &combined & 0.205\\ 
         5 &split & 0.046\\ \hline
    \end{tabular}
    \caption{Timing results for computing the signal and timing results for a short test run. The first column shows the number of parameters being interpolated. The second column show the interpolation method, either the old combined method or the new split method. The third column gives the average timing of running \texttt{photosphere.signal()} a hundred times. These results can be reproduced by running the \href{https://github.com/xpsi-group/xpsi/blob/92bfa8fe8db46a1005b479e8253454e20a0bb2d0/examples/examples_modeling_tutorial/TestRun_AMXP.py} {example}.}
    \label{tab:test_signal}
\end{table}

Lastly, we did an analysis to balance resolution settings \texttt{sqrt\_num\_cells}, \texttt{num\_energies} and \texttt{num\_leaves} between simulation detail and computational time.\footnote{Note here that we did not balance \texttt{num\_rays}, which defines how many rays are in the grid of rays for which light bending is computed (and between which will be interpolated later). Instead we leave this at a `high-resolution' value, which is commonly used in \texttt{X-PSI} analyses: 512. In a future analysis we also aim to balance the value of this resolution parameter.} For each setting, we reduced their values from `high resolution' values until the error in ln-likelihood grew beyond a threshold value of 0.1, which is low enough that at this level we do not find a significant influence on posterior distributions and evidences compared to higher resolution. \Cref{tab:xpsi_settings} provides the X-PSI resolution settings that were obtained. With these settings, a likelihood evaluation takes around 0.1\,s on a single core. This test was done with parameter values close to \largespot{} (which we list in \Cref{tab:parameters}) and a single circular hotspot with angular radius $\zeta = 15\fdg5$. We caution that we have not tested other regions of the parameter space or more complex hotspot configurations.  Elongated hotspots in particular would require more careful treatment, and possibly a finer surface discretisation \citep[see also][]{Choudhury2024b}.

\begin{table*}
    \caption{\label{tab:parameters} Model parameters and values chosen for each \ac{AMP} \scenario. The priors used for sampling are also given, where $U$ means uniform distribution with lower and upper bound and $N(\mu,\sigma)$ means normal distribution with mean and 1$\sigma$ levels. For normal distributions the cut-off is at 5$\sigma$. Unchanged parameters in \smallspot{} compared to \largespot{} are indicated with a dash.}
    \begin{tabular}{lllcc} \hline \hline
    Parameter (Unit) & Description & Prior Density & \multicolumn{2}{c}{Input value}  \\ 
     & & & \largespot{} & \smallspot{} \\ \hline
    $D$ (kpc) & Distance & $N(2.7,0.3)$ & 2.70 & -\\
    $M$ (M$_\odot$) & Mass & $U(1,3)$ & 1.40 & -\\
    $R_\mathrm{eq}$ (km) & Equatorial radius & $U(3R_{\rm G}(1),16)^a$ & 11.0 &-\\
    $\cos i$ & Cosine inclination & $U(0.15, 0.87)$ & 0.770 & 0.174\\
    $f$ (Hz) & Pulsar frequency & fixed & 401 & -\\
    $\phi$ (cycles) & Phase & $U(-0.25,0.75)$ & 0.226 & 0\\
    $\cos\theta$ & Cosine co-latitude & $U(1, -1)$ & 0.985 & -\\
    $\zeta$ (deg) & Angular radius & $U(0, 90)$ & 89.9 & 30.0\\
    $T_\mathrm{seed}$ (keV) & Seed photon temperature & $U(0.5,1.5)$ & 0.529 & 1.28\\
    $T_\mathrm{e}$ (keV) & Electron slab temperature & $U(25,100)$ & 37.3 & 51.1\\
    $\tau$ (-) & Thomson optical depth & $U(0.5,3.5)$ & 1.53 & 2.00\\
    $T_\mathrm{in}$ (keV) & Inner disc temperature & $U(0.01,0.6)$ & 0.168 & -\\
    $R_\mathrm{in}$ (km) & Inner disc radius & $U(R_\mathrm{eq},R_\mathrm{co})^b$ & 30.8 & -\\
    $N_\mathrm{H}$ (\SI{e21}{cm^{-2}}) & ISM column density & $N(1.17,0.2)$ & 1.17 & -\\\hline
    \end{tabular}\\ 
    \vspace{2 mm}
    \begin{flushleft}
    \footnotesize{$^a$ $R_{\rm G}(1)$ is the gravitational radius of $M=1$M$_\odot$. This prior is also modified by the compactness condition, which is detailed in \Cref{sec:prior}.}\\
    \footnotesize{$^b$ $R_{\rm co}$ is a function of $M$ and $f$. See \cref{eq:corotationradius}.}
    \end{flushleft}
\end{table*}

\section{Method}\label{sec:method}
We perform parameter inference with two synthetic datasets, with the aim of testing how well parameters can be recovered for \acp{AMP} with \ac{NICER} data. In \Cref{sec:inference} we first describe parameter inference and how it is applied in this work. In \Cref{sec:scenarios} we then describe the two scenarios, i.e. datasets created by parameter vectors that are both inspired by \ac{J1808}. Lastly, we describe the choices for prior distributions in \Cref{sec:prior}.

\subsection{Parameter inference}\label{sec:inference}
To estimate posterior distributions, we perform nested sampling with \texttt{MultiNest} \citep{Skilling2004, Feroz2009, Buchner2016, Feroz2019}, which is the main nested sampling package used with \ac{X-PSI} to date. Nested sampling is an algorithm which employs iso-likelihood contours to explore parameter space of increasing likelihood to estimate posterior density distributions as well as evidences. In this work we are interested in the posterior distributions, which describe how tightly and accurately parameters $\theta$ of model $M$ are inferred, conditional on a set of data $D$. A posterior distribution is a probability distribution and is expressed with Bayes' theorem: 
\begin{equation}\label{eq:posterior}
P(\theta|D,M)=\frac{P(D|\theta,M)P(\theta|M)}{P(D|M)}.
\end{equation}
Here, $P(D|\theta,M)$ is the likelihood function, which is the probability to obtain $D$ conditional on $\theta$ and $M$, and $P(\theta|M)$ is the prior, which is the probability distribution of $\theta$ without the conditionality on $D$. The evidence $P(D|M)$, also known as marginal likelihood, is notably not conditional on $\theta$, so it is only a normalization factor when computing the posterior distributions. By comparing posteriors to priors we get a sense of how well the data constrains (or `updates') our prior knowledge of the parameters.

We use a Poisson likelihood \citep[as prescribed in][]{Lo2013, Miller2015}, but in contrast to previous work on \acp{RMP} in \ac{X-PSI}, we do not marginalise the likelihood over background parameters \citep{Riley2019}. Instead, we assume the background to be zero. Here, background means any non-pulsed component which is not modelled.  In terms of non-astrophysical background, this choice is justified because we do not expect a significant contribution of counts from sources such as sky background, instrument noise, or space weather \citep{Bogdanov2019a} in the data selection scenario that we are assuming. This is namely a data interval from the peak of a \ac{J1808}-like accretion outburst. In that case we expect the source count rate to be much higher ($\sim$ 250 cnt\,s$^{-1}$, \citealt{Bult2019}) compared to \acp{RMP} (e.g. $\sim$ 0.028 cnt\,s$^{-1}$ for J0740+6620, \citealt{Salmi2022}), for which non-astrophysical background is thus much more significant. We verified this assumption with \ac{NICER} data at the peak of the \ac{J1808} outbursts in 2019 and 2022 and found that the estimates for instrumental background count rate go up to 2 per cent of the total counts, but typical values are lower than 1 per cent \citep{Bult2020, Illiano2023}. While this contribution is non-zero, it is small, and we leave the analysis of any bias caused by this assumption for future work. In terms of astrophysical background, there is a non-pulsed component relevant for \acp{AMP}: the accretion disc. Rather than treating this as an unmodelled component to be marginalized over \citep[as in e.g.][]{Kini2023}, we instead model this component (see section \Cref{sec:disc}), which allows us to infer disc parameters.
 
\subsection{Synthetic AMP scenarios}\label{sec:scenarios}
This section introduces the two \ac{AMP} \scenarios, i.e. two different simulated datasets. These are produced by inserting different model parameters to the same model to generate data. \Cref{tab:parameters} provides an overview of the parameters in both \scenarios{} and the prior distributions. \Cref{fig:pulse_diagnostics} visualises the pulse profiles by showing the pulses at various energies, the fractional amplitudes and phases as a function of energy, and the decomposed spectra. For all parameters and their prior distributions, an attempt was made to base them on the literature for \ac{J1808}, such that both \scenarios{} -- while different -- are representative for analysis of \ac{J1808} to be done in future work. For each \scenario, we run the analysis three times using data with different Poisson noise realizations, such that it becomes clearer if biases are systematic. The synthetic data is generated with 132 ks of exposure time in both \scenarios, and we give more details about this choice in \Cref{sec:largespot}.

First, let us list parameters which are the same between the two \scenarios. For $M$ and $R_{\rm eq}$, we used typical \ac{NS} values commonly used in the literature for \ac{J1808} \citep[e.g.][]{Kajava2011}. For $D$, we used the latest distance estimate of 2.7\,kpc from \cite{Galloway2024}. The value for $N_{\rm H}$ was computed with the HEASoft nH tool \citep{HEASoft} and the Hi4Pi map \citep{bekhti2016}. The pulsar frequency $f$ is well measured \citep[e.g.][]{Bult2020} and is a fixed parameter during sampling. \Cref{fig:comparison_pf} shows the simulated \ac{NICER} data of the two \scenarios\ in the third panels from the top. Below, we discuss the qualities unique to each \scenario.

\newcommand{\plotwidth}{0.49\textwidth}

\begin{figure*} 
    \centering
    \begin{minipage}[b]{\plotwidth}
        \centering
        \includegraphics[width=\textwidth]{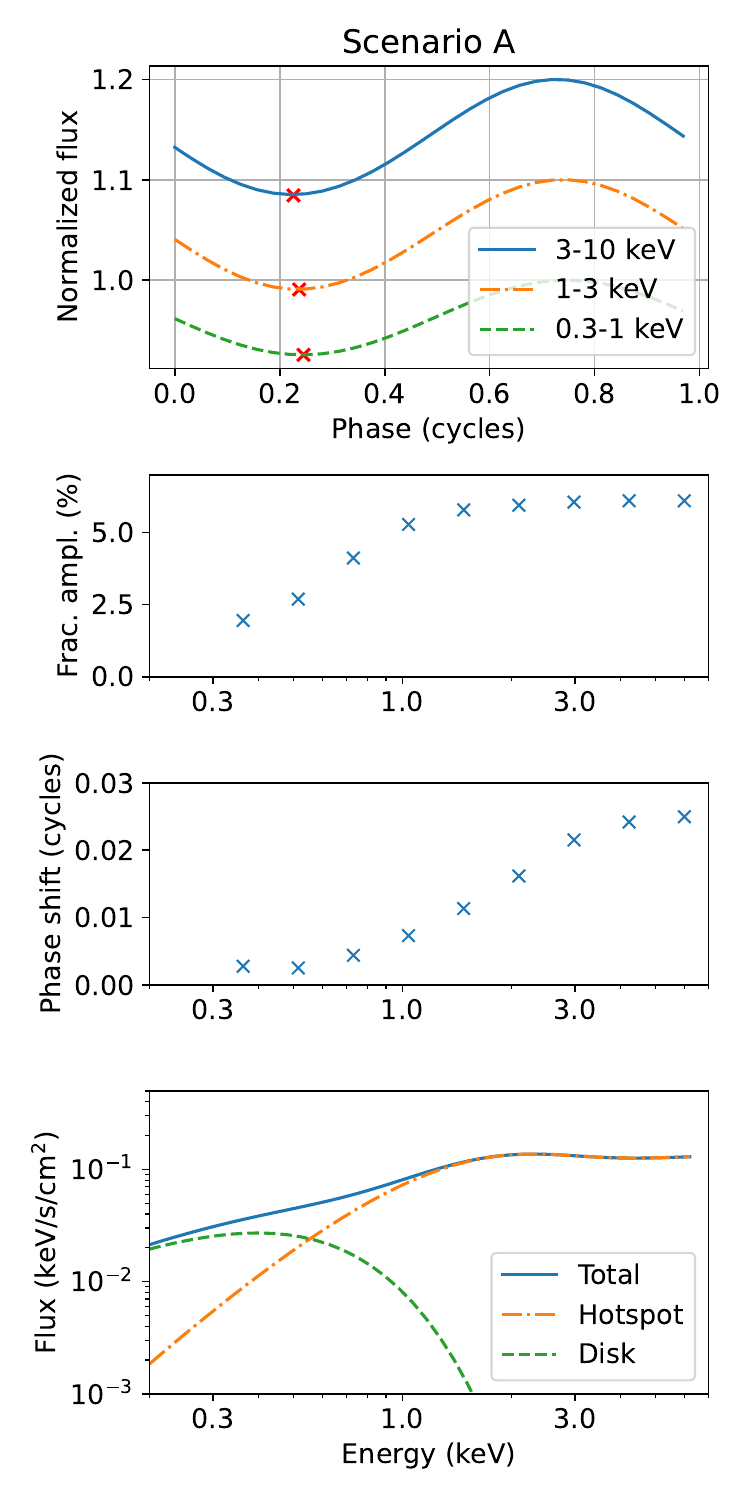}
    \end{minipage}
    \hfill
    \begin{minipage}[b]{\plotwidth}
        \centering
        \includegraphics[width=\textwidth]{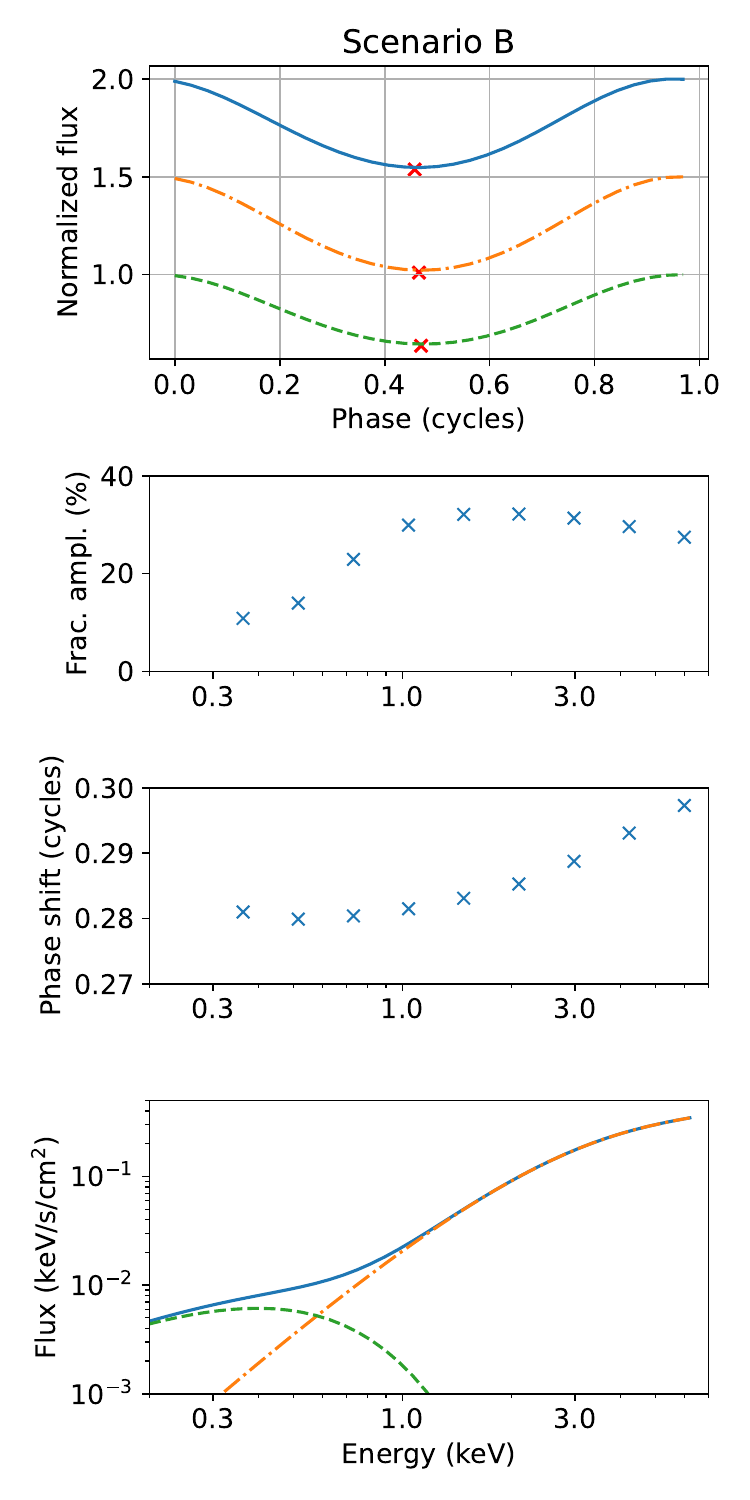}
    \end{minipage}
        \caption{Various properties of the pulse profiles of \largespot{} on the left panels and \smallspot{} on the right panels. The top plot displays the normalized pulse profile, where the flux has been rebinned into low, middle and high energy: 0.3--1 (dashed line), 1--3 (dash-dotted line), and 3--10~keV (solid line). These pulse profiles have been arbitrarily shifted upwards for clarity. The red crosses guide the eye to track the relative phase shift by marking the minima of sine functions that were fitted to each pulse profile. The second and third plots show respectively the fractional amplitude and phase shift as a function of energy. These were computed after rebinning the emission into logarithmically spaced energy bins. The bottom plot shows the total phase-averaged spectrum (solid line) as well as the decomposition into the two contributing components: the hotspot (dash-dotted line) and the disk (dashed line).}
    \label{fig:pulse_diagnostics}
\end{figure*}

\begin{figure*} 
    \centering
    \begin{minipage}[b]{\plotwidth}
        \centering
        \includegraphics[width=\textwidth]{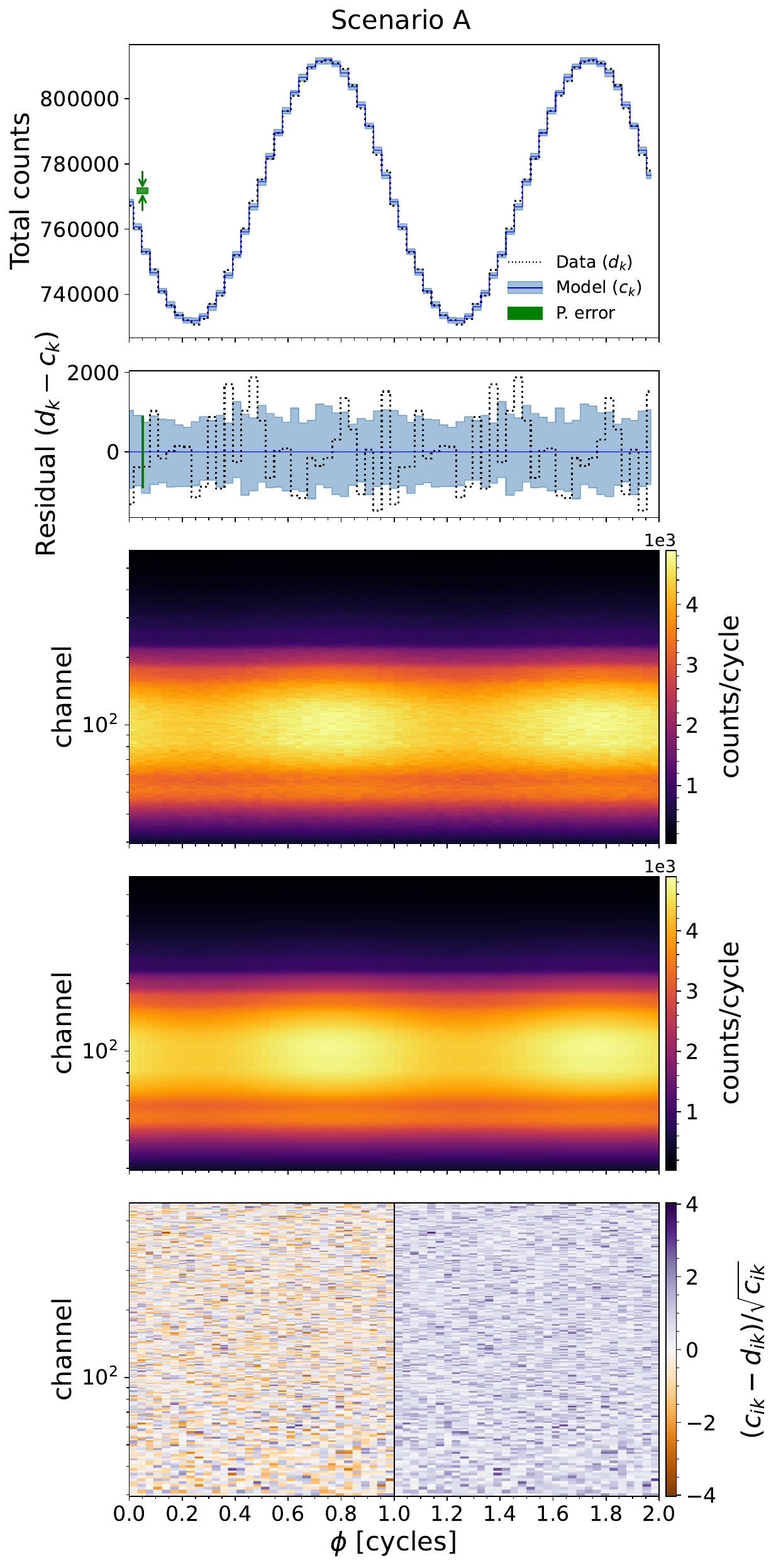}
    \end{minipage}
    \hfill
    \begin{minipage}[b]{\plotwidth}
        \centering
        \includegraphics[width=\textwidth]{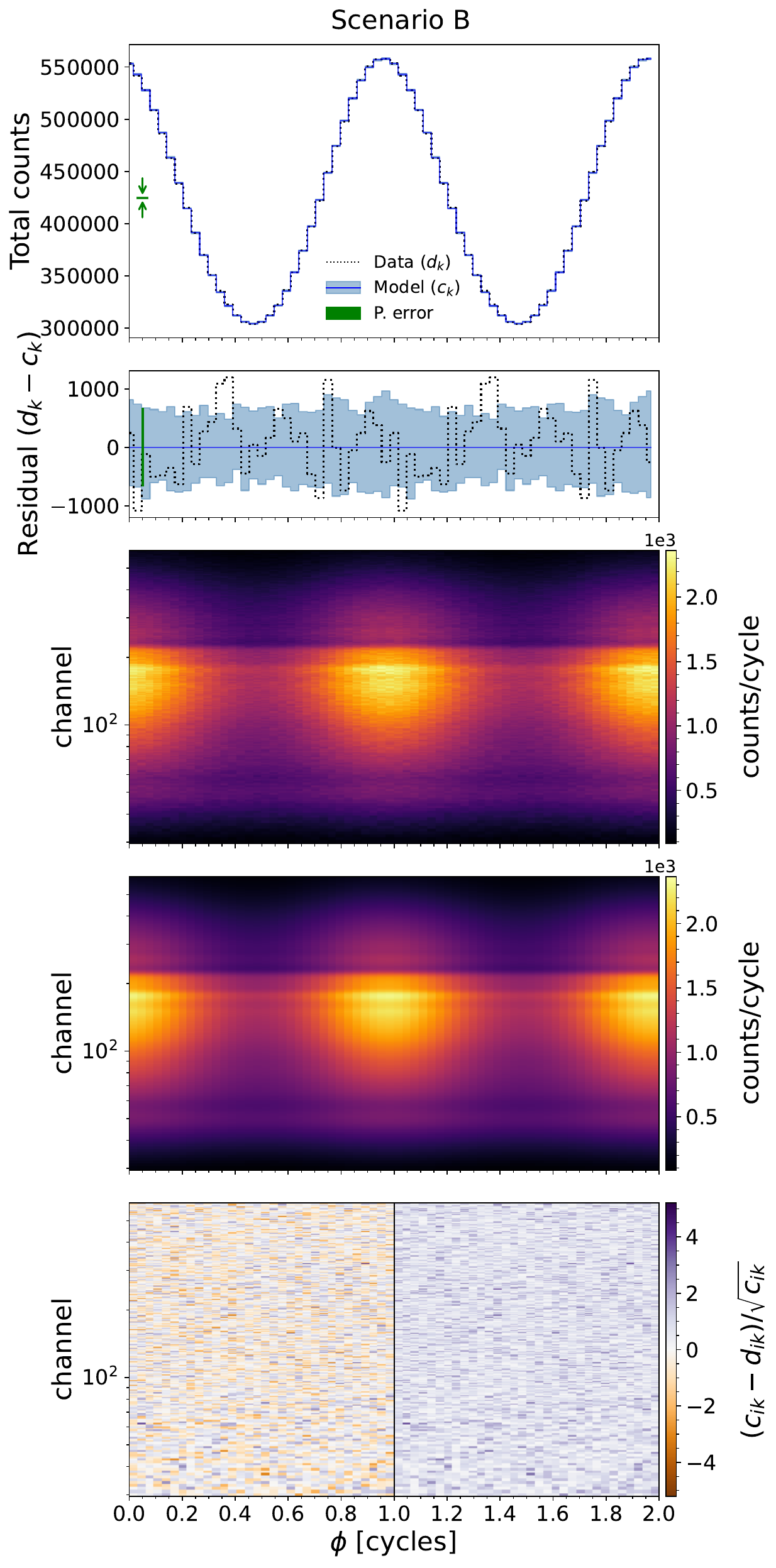}
    \end{minipage}
        \caption{The top panels show a comparison between the bolometric pulse data ($d_k$) and bolometric posterior expected pulse from 100 equally weighted posterior samples ($c_k$). Here, the index $k$ corresponds to phase bins. The blue band corresponds to the 16 to 84 per cent credible interval. The $\pm1\sigma$ Poisson error at the mean of the pulse is shown in green on the left. In the second panels the black dotted line show shows the bolometric residual $d_k-c_k$. Here, the blue band shows the same credible intervals around the model, centred at zero. The same Poisson error is shown again for reference in green. The third panels show data realizations with seed 1 of synthetic \scenarios{} A (left) and B (right). The colour corresponds to the total \ac{NICER} counts per cycle per phase-energy bin in a total exposure time of 132~ks. The x-axis corresponds to the phase and the y-axis corresponds to the \ac{NICER} energy channels, where channel 100 corresponds roughly to 1~keV. The fourth panels are similar, but show the posterior expected pulse profiles from 100 samples. Finally, the bottom panels show the normalized residuals of the data and model where $c_{{ik}}$ is the posterior expected pulse profile and $d_{{ik}}$ is the data. Here, the index $i$ corresponds to energy channels. The plots with seeds 2 and 3 are not included because they are visually almost indistinguishable from seed 1.}
    \label{fig:comparison_pf}
\end{figure*}

\subsubsection{\largespot{}: fit of 2019 NICER data of J1808}\label{sec:largespot}
This \scenario{} is a synthetic dataset that fits closely to real \ac{NICER} data from around the peak of the 2019 outburst of \ac{J1808}.

The \ac{NICER} data imitated here is a selection from 10 up to and including 20 August 2019, which includes 132 ks of exposure time. This exposure time is much shorter than typical for \acp{RMP}, but because \acp{AMP} such as \ac{J1808} are very bright during peak outburst, \ac{NICER} detected $2.5\times 10^7$ counts. There is no phase shift in the pulse profile in this selection except for some minor phase shift in the first two days \citep[see][]{Bult2020}. The pulse profile shapes are stable throughout this selection.

The fitting was done using the same sampling method as presented in \Cref{sec:inference}. However, here fewer parameters were sampled because $M, R_{\rm eq}, D$, and $N_{\rm H}$ were kept fixed, and the lower limit in $i$ was $\sim$40\degr\ \citep[which was also used by][]{Salmi2018}. An earlier fit with $M$, $R_{\rm eq}$, $D$ and $N_{\rm H}$ free led to very large $M$, small $R_{\rm eq}$ and $D$ near the edge of the prior, far removed from what we would expect for typical \acp{NS} and \ac{J1808} in particular, causing us to fix those parameters. We plan to analyse the real data in more detail in a future paper.

A best-fit is found with a large hotspot covering just under half of the star. While the fit accounts well for the pulsed component, some phase independent residuals were left around 1~keV, which can perhaps be attributed to broadened Fe or Ne lines \citep[also noted by][]{Sharma2023}, but are not modelled here. The physics of 1~keV lines of X-ray binaries have also recently been investigated by \cite{Chakraborty2024}.

\subsubsection{\smallspot{}: polarized radiation}
While \largespot{} fits reasonably well with the \ac{J1808} data, one drawback is that such a large circular hotspot, in combination with both small inclination and co-latitude, does not predict a detectable degree of polarized radiation. This occurs because polarization is averaged out across the large hotspot surface, and some observed photons are always emitted at small angles relative to the surface normal, where polarization is minimal. However, recently, \cite{Papitto2025} detected polarized radiation from an \ac{AMP} with {\it IXPE} in a campaign that included joint \ac{NICER} observations.

Given this context, we used \textsc{ixpeobssim} \citep{Baldini2022} to generate another synthetic dataset to analyse. This dataset does not fit closely to the \ac{NICER} data of \ac{J1808}, but represents an example of an \ac{AMP} which produces polarized radiation.

For this \scenario, we modify \largespot{} by giving it a smaller $\zeta$ and larger $i$ so that more polarized radiation is detected. We also modified $T_{\rm seed}, T_{\rm e}$ and $\tau$, until we found values that resulted in more than $\sim$2 per cent  polarization degree. Although the aim was to keep the higher X-ray counts at high energy roughly similar, in this scenario they are around a factor of two higher. We note, however, that now simulating with the same exposure time yields substantially fewer \ac{NICER} counts: $\sim$1.4$\times 10^7$.

\subsection{Prior densities}\label{sec:prior}

Both \scenarios{} share the same prior densities which are inspired by previous studies and our early stage analyses of \ac{NICER} data of \ac{J1808}. An overview of the priors is given in \Cref{tab:parameters}. 

For mass and radius priors, we follow previous X-PSI conventions \citep[e.g.][]{Riley2019}, namely wide uniform ($U$) distributions for the \ac{NS} mass and the radius with $U(1,3) {\rm M}_\odot$ and $U(3R_{\rm G}(1), 16)$ km respectively. Here $R_{\rm G}(M) = GM/c^2$ is the gravitational radius (and 3$R_{\rm G}(1) = 3G{\rm M}_\odot/c^2 \approx$ 4.4 km). The prior is further modified by setting the compactness limit at $R_{\rm pole}/R_{\rm G}(M)>2.9$ \citep[see e.g.][]{Gandolfi2012}, where $R_{\rm pole}$ is the polar radius. For the distance prior we use a Gaussian distance estimate $2.7 \pm 0.3$ kpc with a cut-off at $5\sigma$, which is inspired by the latest distance estimate from \citet{Galloway2024}. For $N_{\rm H}$ we assume a Gaussian prior with $(1.17\pm0.2)\times 10^{21}\rm{cm}^{-2}$ and a cut-off at $5\sigma$. As a prior density for the cosine inclination we use $U(0.15, 0.87)$, where the lower limit excludes close to edge-on angles due to the lack of X-ray eclipses \citep{Chakrabarty1998} and the upper limit was increased from 0.77 rad (40\degr, used when sampling the real data in \Cref{sec:largespot}) to 0.87 rad (30\degr). This allows us to explore smaller inclinations than the input parameter, which is $\sim$40\degr\ in \largespot. The hotspot is allowed to have any phase but the prior distribution is chosen to not have an edge at zero, $U(-0.25, 0.75)$ cycles. In addition we allow any hotspot co-latitude: $U(0\degr, 180\degr)$.\footnote{If the hotspot would be on the other side of the star from the observer, one could argue that a second hotspot should become visible, but a second spot has not been implemented here. Limiting the co-latitude to 90\degr\ would alleviate this. We have not implemented this change, but because we end up inferring a small co-latitude, this upper limit does not come into play.} The angular radius is such that the hotspot is allowed to maximally cover half of the star: $U(0\degr, 90\degr)$. For the atmosphere parameters $T_\mathrm{seed}, T_\mathrm{e}$ and $\tau$, we uniformly cover the full range of the pre-computed data. The disc inner radius $R_\mathrm{in}$ is allowed to vary between $R_{\rm eq}$ and the co-rotation radius 
\begin{equation}\label{eq:corotationradius}
    R_\mathrm{co} = (GM/4f^2\pi^2)^{1/3},
\end{equation}
because outside of that radius, accretion would be in the propeller regime. Whether accretion continues in that regime is a subject of debate \citep[see e.g.][]{Rappaport2004, Romanova2005, Kluzniak2007, DiSalvo2022}. However, because we will -- when we do analysis on real data -- select data from near the peak of the outburst, we expect the accretion rate to be too high for the system to be near the propeller regime. Lastly, $T_\mathrm{in}$ has a broad uniform prior distribution of $U(0.01,0.6)$~keV.

\section{Results}\label{sec:results}
This section shows the results of the parameter inference runs, subdivided into results for \largespot{} and \smallspot. We examine the parameter recovery in terms of precision and accuracy, and especially look for systematic biases present in the posterior distributions (i.e. present for all seeds). For all parameter inference runs presented in this work we set the sampling efficiency in \ac{X-PSI} to 0.1. We also always use 1000 live points. In test runs we found that this modest number of live points was sufficient, because there were no significant changes to the posteriors distributions beyond this number. 

For both scenarios, the top panels in \Cref{fig:comparison_pf} show the bolometric data and posterior expected pulses. The 16 to 84 percent credible intervals of those are pulses are indicated by a blue band. The second panels show the bolometric residuals, which indicate that the model fits the data well, with only some remainder due to  $\pm1\sigma$ Poisson noise ($\sqrt{\rm counts}$). This panel also shows that the $\pm1\sigma$ Poisson error is similar in size to the average 16 to 84 per cent credible interval for the posterior expected pulse. The third and fourth panels show the data and posterior expected pulse profiles respectively and the bottom panel shows the residual between the data and the posterior expected pulse profiles. There is no structure in these residuals, which indicates that the model fits the data well, with only Poisson noise unaccounted for.

\subsection{Inference results in \largespot}
\begin{figure*}
    \centering
    \includegraphics[width=1.0\textwidth]{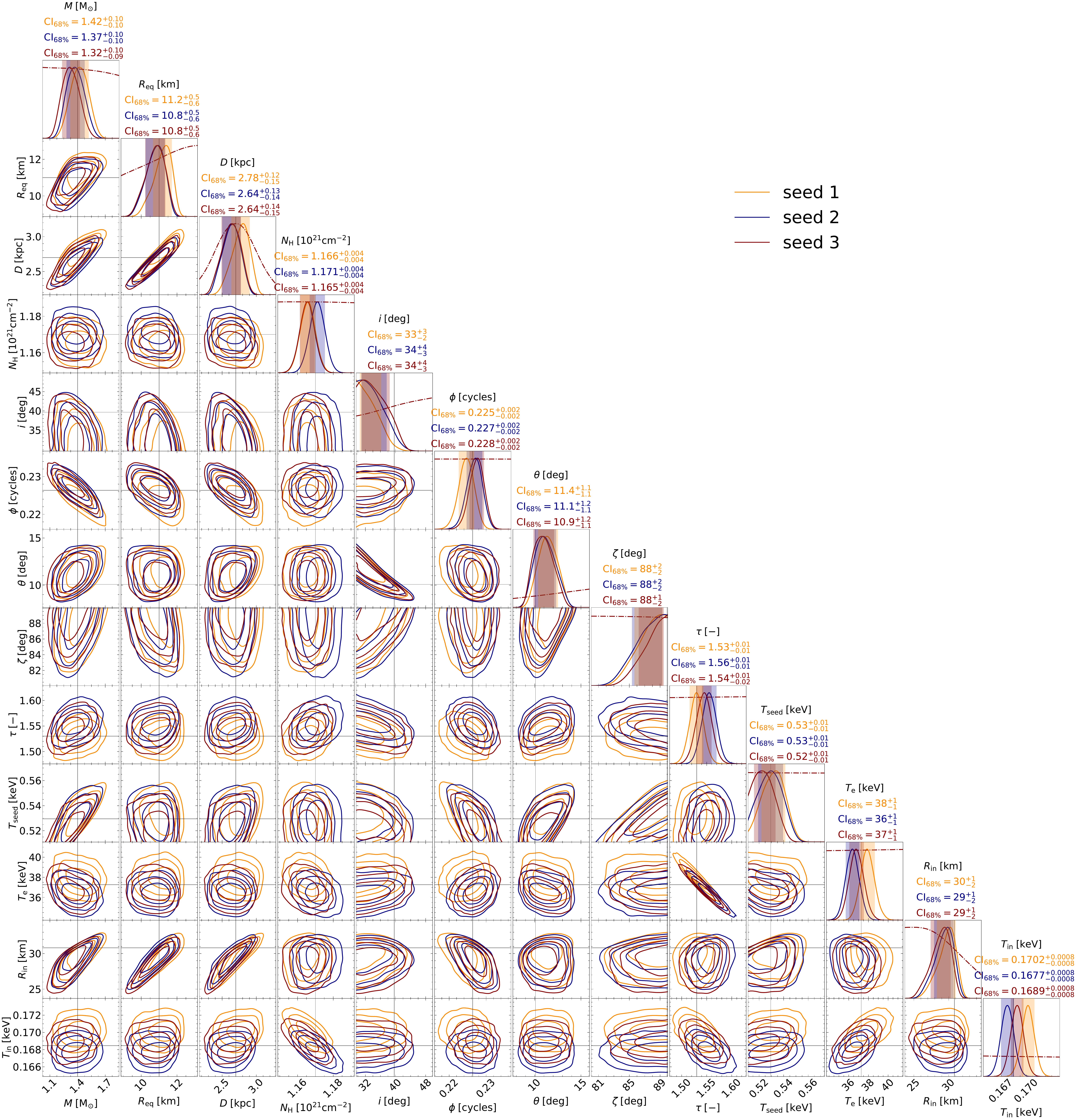}
    \caption{Cornerplot of all parameters in \largespot. The three colours correspond to the analyses with three different noise realizations for the synthetic data. Solid curves are the \ac{1D} and \ac{2D} marginalised posteriors, dashed curves are the priors, and black lines are the truths. Contours in the \ac{2D} plots are the 68, 95 and 99.7 per cent credible levels. The median of the posterior distributions are quoted above, with \ac{1D} credible intervals quoted at the 68 per cent level. These credible intervals are also visualised by the coloured intervals on subplots along the diagonal.}
    \label{fig:cornerplot_large_r}
\end{figure*}

\Cref{fig:cornerplot_large_r} shows the posterior distributions resulting from the parameter inference. Many parameters are recovered with high precision. To highlight the two main parameters of interest: $M$ is recovered at the 1$\sigma$ level at $\sim\pm$7 per cent precision ($\pm 0.1$M$_\odot$) and $R_{\rm eq}$ at $\sim\pm$6 per cent precision ($\pm 0.6$ km). In addition, there are no systematic biases for these two parameters, so it can be concluded that these are recovered robustly in this \scenario.

For most parameters we find that the input value is within the 1$\sigma$ credible interval of the posterior, which indicates the parameters are recovered accurately in most cases. The exceptions are $i, \theta, \zeta$ and $R_{\rm in}$. In the case of $\zeta$, this bias in the posterior can be explained by the true value being very close to the upper bound of a flat prior, combined with lower values not being strongly disfavoured under the constraints given by the data. $R_{\rm in}$ is a similar case, but here the upper bound of the prior is limited by the $R_{\rm co}$ which scales with $M^{1/3}$ (see \cref{eq:corotationradius} and also the prior of $R_{\rm in}$ in Fig.~\ref{fig:cornerplot_large_r}) which implies lower $R_{\rm in}$ has higher prior probability.

We believe that the biases in $i$ and $\theta$ result from the bias in $\zeta$ and $R_{\rm in}$, which are in turn caused by the prior. This is because of the effect these parameters have on the pulse profile in this regime. Namely, we find a decreased count rate and increased pulse fraction when we plot pulse profiles with the median posterior value, rather than the input value, for $\zeta$ and $R_{\rm in}$, while keeping all other parameters at their input values. Correspondingly, we also find that we can compensate for this change in pulse profile by shifting $i$ and $\theta$, which show slight mutual degeneracy, in the direction in which they are biased in Fig.~\ref{fig:cornerplot_large_r} (i.e. reduce $i$ and increase $\theta$). 

\subsection{Inference results in \smallspot}
\begin{figure*}
    \centering
    \includegraphics[width=1.0\textwidth]{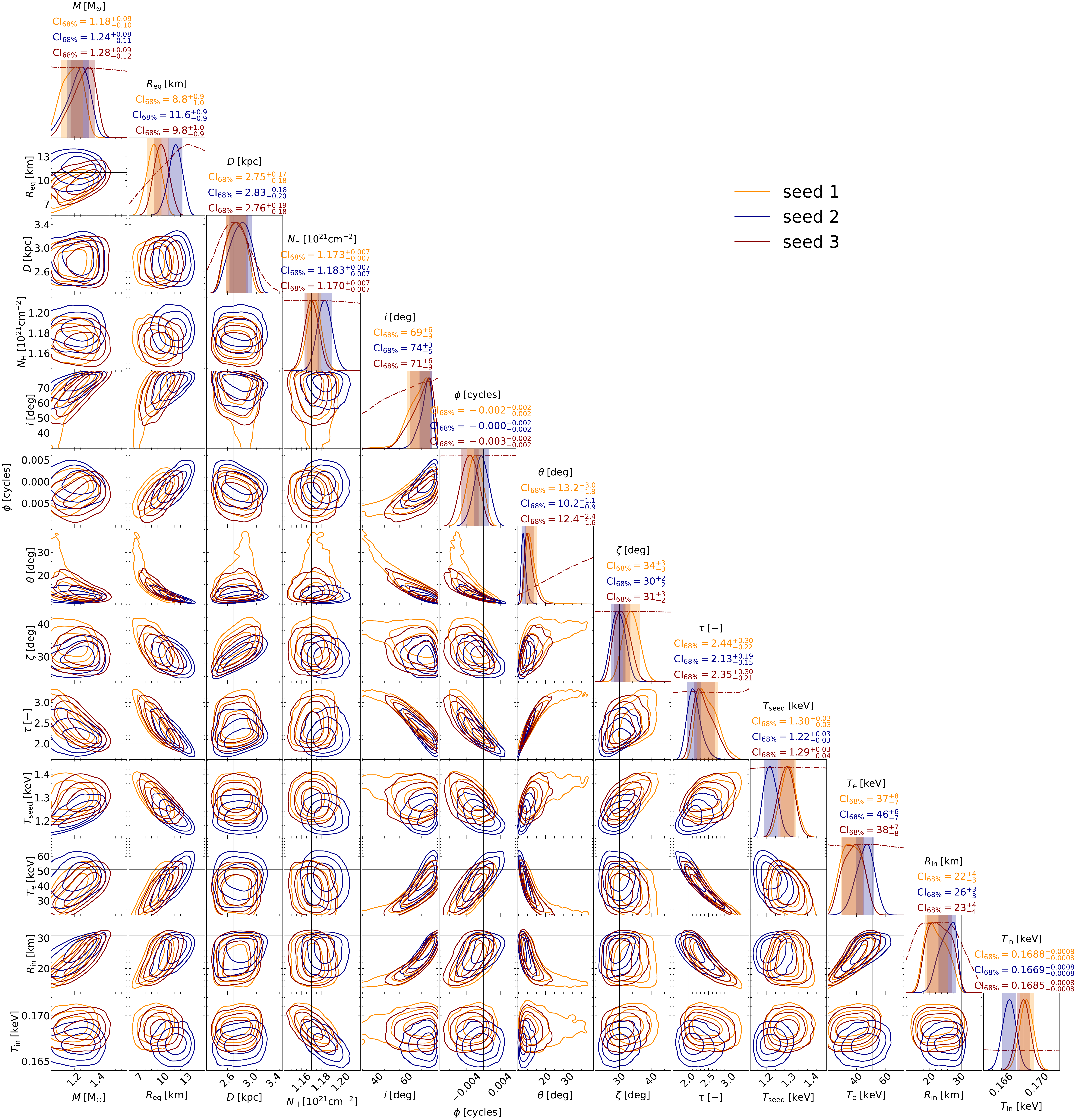}
    \caption{Cornerplot of all parameters in \smallspot. Solid curves, dashed curves and black lines are the posteriors, priors and truths respectively. Contours are the 68, 95 and 99.7 per cent credible levels.}
    \label{fig:cornerplot_small_r}
\end{figure*}

\Cref{fig:cornerplot_small_r}, which summarizes the results for \smallspot, shows slightly worse performance in terms of parameter recovery. While many parameters are still accurately recovered, there are now some significant systematic biases in a subset of the parameters. These biases are exacerbated by more (pronounced) degeneracies present between some model parameters. We also see that some of the posteriors of these parameters (on the diagonal) are widened.

To highlight again the main parameters of interest: $M$ is not recovered as accurately, with the true value being systematically higher than the 1$\sigma$ credible interval in all cases, but the posterior still has a similar precision of around $\sim\pm 8$ per cent ($\pm 0.1 M_\odot$). In two out of three cases, $R_{\rm eq}$ is also not recovered at 1$\sigma$ level, while the posterior has also slightly widened to  $\sim\pm 9$ per cent precision ($\pm 0.9$ km).

As in \largespot, $i$ and $\theta$ are degenerate with each other, as well as $\tau$ and $T_{\rm e}$. However, what is different now is that these four parameters are all degenerate with each other and some also show degeneracy with $M$, $R_{\rm eq}$ and $R_{\rm in}$. The degeneracy between $\cos i$ and $R_{\rm in}$ can be explained by the fact that they are both proportional to $F_{\rm disc}$ (see \Cref{eq:disc}). However, the degeneracy is only noticeable in \smallspot{} because the disc is viewed more edge on. This means a small bias in $\cos i$ leads to a relatively larger effect on $F_{\rm disc}$, which implies a more pronounced compensation in $R_{\rm in}$ to keep $F_{\rm disc}$ constant, making the degeneracy more apparent.

A possible origin for the biases could be $R_{\rm in}$, which again is near the upper bound of the prior. We found that rerunning the analysis, but with a fixed $R_{\rm in}$, removes degeneracies between $M$ and other parameters, as well as tightening the posterior distribution of $M$ and removing the bias. It also appears to improve the bias on $i$, leaving both true values within 1$\sigma$. This hints at some degeneracy between disc and star flux. We discuss the implications in \Cref{sec:discussion}. We also found a similar but less pronounced effect with a fixed $\theta$, which improves but does not remove the bias on $i$ (true value within 2$\sigma$) and $M$ (true value within 1$\sigma$).

This \scenario{} serves as an example of how bias in one parameter can lead to a bias in another if degeneracy is present, making the parameter recovery less robust. This effect is inherent to a combination of the model and the parameter regime it is in, as well as the dataset. However, complementary data such as polarization information could alleviate the degeneracy in some parameter regimes.

\section{Discussion}\label{sec:discussion}

\subsection{Summary of results}
In \Cref{sec:results}, we analyzed synthetic datasets of two \ac{AMP} scenarios. In \largespot{}, which is characterized by its large hotspot, we found generally good recovery of most model parameters, including the main parameters of interest $M$ and $R_{\rm eq}$. However, we found some systematic bias in $R_{\rm in}, \zeta, i$ and  $\theta$. We proposed that the bias in $R_{\rm in}$ and $\zeta$ could be due to the input values being close to the upper bound of the prior, and that $i$ and $\theta$ are biased as a result of a compensation. In \smallspot{}, we found slightly worse performance in terms of parameter recovery, including more biases and widened posteriors. We propose that this is the result of more degeneracies in this parameter regime: between $M, R_{\rm eq}, i, \theta, \tau, T_{\rm e}$ and $R_{\rm in}$. We find a systematic bias on $M$, such that the true value lies outside (above) the 1$\sigma$ level of the posterior in all three noise realizations. We also find that rerunning the analysis with a fixed $R_{\rm in}$ drastically improves the recovery of $M$. This hints at degeneracy between the disc and the star, which we discuss in \Cref{sec:innerdisc}. We note that the reduced counts in this \scenario{} could also (partially) account for widening of the posteriors.

\subsection{Comparison with other work}
Let us first focus on the inference of $M$ and $R_{\rm eq}$. The precision of the inference with simulated \ac{NICER} data is similar to the results of \cite{Salmi2018}, when they inferred parameters with synthetic {\it RXTE} data. Assuming no constraints on $i$ and $\theta$ from polarization data from {\it IXPE}, they found a 68 per cent credible interval of 0.20 M$_\odot$ for $M$ (relative error in the value for $M$ of 13 per cent), where we also find 0.20 M$_\odot$ (14 per cent) and 0.20 M$_\odot$ (16 per cent) when averaging the results from the three seeds in \largespot{} and \smallspot{} respectively. Also, they found an interval of 1.7 km (14 per cent) compared to our on average 1.1 km (10 per cent) and 1.9 km (19 per cent) in \largespot{} and \smallspot{}, respectively.

The number of counts assumed by \cite{Salmi2018} was similar: $4 \times 10^7$ counts for {\it RXTE}, compared to 2.5 (1.4) $\times 10^7$ \ac{NICER} counts in \largespot{} (\smallspot). This gives some indication that the constraints achievable with \ac{NICER} can be similar to {\it RXTE}, even though the latter probed a higher energy part of the spectrum.\footnote{The {\it RXTE} band (2 to 60 keV) is a higher energy part of the spectrum where the beaming function from the hotspot is different \citep[see e.g. Fig. 1 in][]{Bobrikova2023} and the reflection component off the disc is important, while the black body component of the disc is not.} However, caution should be taken with this conclusion because there are substantial differences in the \ac{NS} models: they used a different spectral and beaming model for the radiation from the hotspot, and their hotspot is (much) smaller, with an angular radius of $15\fdg5$.

Then, let us compare the inference of $i$ and $\theta$. In our case we find tight constraints, but we also find that these parameters are mutually degenerate and that their inference was biased. \cite{Salmi2018} found comparable constraints with synthetic {\it RXTE} data. For $i$ they found a 68 per cent credible interval of 12\degr, compared to our on average 6\degr\ (13\degr) in \largespot{} (\smallspot). For $\theta$ they find an interval of 3\degr, compared to our average of 2\degr\ (3\degr) in \largespot{} (\smallspot). 

If the polarization degree is sufficient, complementary constraints on $i$ and $\theta$ could be obtained through joint detection with {\it IXPE}. Geometry constraints might improve inference of other parameters as a result, for example in \smallspot{} we found that there is a bias in $M$ which could be improved by fixing $\theta$.
We can get an indication of the polarization constraints by looking at the results for model 3b by \cite{Bobrikova2023}, which in terms of $i (60\degr)$, $\theta (20\degr)$ and $\tau (1.6)$ is not identical but most comparable to our \smallspot. They found a 68 per cent credible interval on $i$ of 25\degr, while we find on average 13\degr{} and some hint of a systematic bias of $-9\degr$. They also found a credible interval on $\theta$ of 7\degr{} while we find 3\degr, and no clear indication of a systematic bias. While sizes of the credible intervals are somewhat larger for polarization data, it is still possible that joint analysis would yield tighter and more accurate constraints. Because these data independently measure $i$ and $\theta$, the overall effect of any systematic bias present in only one of the two could be reduced. We also note some differences in their analysis: they fix mass and radius and only fit normalized Stokes $q$ and $u$ parameters. 

While we have not yet explored joint \ac{NICER} and {\it IXPE} analysis, we expect that the combination of timing and polarization data would lead to tighter and more accurate constraints. We leave the exploration of joint analysis for future work.

\subsection{Inner disc radius}\label{sec:innerdisc}

A parameter of importance appears to be $R_{\rm in}$. In both \scenarios{} the recovery of this parameter is biased, which we believe is due to the true value being close to the upper bound of the prior. This upper bound is determined by $R_{\rm co}$, which scales as $M^{1/3}$, so the prior probability of sampling high values of $R_{\rm in}$ is reduced because that requires that a sufficiently high mass is also sampled. In both \scenarios{} we find hints that bias in $R_{\rm in}$ contributes to biases in some other parameters as well. It is clear that fixing $R_{\rm in}$ resolves the bias in $M$. We conclude that there could be some degeneracy between the flux from the star and the disc. 

Given the importance of $R_{\rm in}$, it is clear that accurate modelling of the disc is crucial for accurate parameter inference. While $R_{\rm in}$ represents the inner disc radius apparent from the pulse profile, it must be noted that the disc model we use is a simple model which does not include some important physical effects that could offset $R_{\rm in}$ from the actual inner disc truncation radius $R_{\rm t}$. We discuss improvements to the disc model in \Cref{sec:caveats}. In addition, it has been proposed that magnetic field--disc coupling could lead to a dark inner region of the disc, such that $R_{\rm t} < R_{\rm in}$ \citep[e.g.][]{Kajava2011}.

We also stress the importance of the prior on this parameter. Even under the assumption that the disc model perfectly represents reality, the closeness of the true value of $R_{\rm in}$ to the upper bound creates a bias. We tested the effect of removing $R_{\rm co}$ as an upper bound and replacing it with an upper bound of 50 km, which encompasses 31 km for the $R_{\rm co}$ of \ac{J1808}. This widens the (spread of) posteriors making the inferred value for $R_{\rm in}$ potentially less reliable and (correspondingly in the case of \smallspot{}) the value for $M$ as well. We conclude that in an analyses of real data the prior on the disc should be physically well motivated, because otherwise there is a risk of biasing the inference of $M$.

Constraining $R_{\rm in}$ through other methods would be helpful in this effort. One avenue would be constraining $R_{\rm in}$ through the shape of the broadened iron line \citep{Papitto2009, Cackett2009, Chakraborty2024, Ludlam2024}. Another avenue could be to get an estimate of the accretion rate along with the star's magnetic field, which links to $R_{\rm m}$ \citep{D'Angelo2010, Kulkarni2013}. Estimation of the accretion rate for \ac{J1808} has been recently discussed by \citet{Illiano2023}. The possibility of non-conservative mass transfer also needs to be contended with as it affects the estimation of the B-field \citep[see e.g.][]{Marino2019, Bult2021, Ng2021}.

\subsection{Caveats and possible improvements}\label{sec:caveats}

Our \ac{AMP} model includes several assumptions and simplifications that could well affect the results.
A first issue is the disc model \texttt{diskbb}, which currently only consists of an additive flux contribution to the total flux. One improvement would be to include disc geometry (ideally in three dimensions), such that light rays would be attenuated if a light ray had to traverse the disc \citep[this was explored e.g. by][]{Poutanen2009, Ibragimov2009}. Furthermore, as noted by \cite{Done2007}, several pieces of physics are currently missing from the disc model. General relativistic light bending \citep[e.g.][and references therein]{Loktev2022} is currently missing for the disc emission, but could be significant, especially for small $R_{\rm in}$. In addition, spectral hardening \citep{Shimura1995}, where the interplay between the scattering and absorption opacities in the disc leads to a higher observed colour temperature, is not included and could also bias $R_{\rm in}$ \citep{Salvesen2013}. In addition, we do not include (line) reflection and reprocessing of radiation from the hotspot in the inner regions of the disc. This was studied for example by \cite{Wilkinson2011} and \cite{Garcia2013} who present a grid of models.

A second point is the modelling of the hotspots. In this work, we assume that there is only one hotspot due to computational limitations. However, even if a pulse profile is close to sinusoidal, one cannot a priori rule out that it is produced by a \ac{NS} with two visible hotspots. For example Fig.~6 of \citet[][]{Poutanen2006b} demonstrates that a neutron star with two hotspots always visible can be close to sinusoidal at e.g. $i=60^\circ$ and small $\theta$. We conclude by recommending a two hotspot approach for future work.

We also note that the hotspot geometry used here was circular, but \ac{3D} \ac{MHD} simulations of magnetospheric accretion are showing more complex geometries. These have been done by for example \cite{Romanova2004} and \cite{Kulkarni2013} and more recently including general relativity by \cite{Das2024}. For very small $\theta$ (magnetic obliquity, or misalignment between magnetic moment and \ac{NS} rotational axis) the hotspot forms a ring shape, which transforms into a crescent shape when $\theta$ increases. Finally, as the $\theta$ reaches 90\degr, the hotspots become bar shaped. We note that there exist \ac{X-PSI} model classes that could potentially capture these geometries well. They consist of a radiating circle overlapped by another masking circle (CST, EST and PST in e.g. \citealt{Riley2019} or Fig.~1 in \citealt{Vinciguerra2023}.) to form crescents or rings. As a possible alternative to \ac{X-PSI} model classes, analytical formulae of hotspot shapes have also been developed by \cite{Kulkarni2013}. These depend on three parameters: $\theta$, normalized co-rotation radius $R_{\rm co}/R_*$, where $R_*$ is the radius of the \ac{NS}, and normalized magnetospheric radius $R_{\rm m}/R_*$. This parameterization could provide a step up from a circle, but is restricted to the assumption that the magnetic field is dipolar. One major hurdle in implementing this parameterization from \cite{Kulkarni2013} for the purpose of parameter inference would be that these hotspots feature Gaussian flux distributions. This implies stepping away from homogeneous atmosphere parameters which would represent a major increase in computational cost, as described in \Cref{sec:optimisation}.

Besides the hotspot, there is also the question of emission from the rest of the \ac{NS} surface. We have done small exploratory runs with fewer livepoints (192) with a synthetic dataset that includes an ``elsewhere'' component. This is a uniform blackbody contribution from the rest of the \ac{NS} surface. In our test case we used an injected value of 0.5 keV with a very broad uniform prior, and we found in those runs that the elsewhere and the accretion disk both contribute at low energy (around 2 keV and below). In these exploratory runs we saw significant differences in their relative contribution, and we also found generally that posterior probability distributions widened. 

While these runs are small and the results are only preliminary, we can at least say that in real data we expect the pulse fraction to be low at lower energies (see e.g. Fig.~2 in \citealt{Bult2020}). This is consistent with the dominant contribution being of the accretion disk, because we expect the elsewhere component to produce still some pulsations at lower energies.

Another issue is the accretion environment, which includes many physical effects that are not modelled here. For example, recently \cite{Ahlberg2024} showed that scattering in the accretion funnel has a significant effect on the pulse profiles of \acp{AMP}, and this depends on accretion rate.

Lastly, we note that our model is agnostic of the values of the magnetic field and accretion rate. However, it should be possible to define a model based on the magnetic field properties, on which other parameters depend in a self-consistent manner. For example, $R_{\rm m}$ (which corresponds to $R_{\rm in}$) depends on both magnetic field and accretion rate. In addition, because the accreted material follows magnetic field lines as it is accreted from the disc, the hotspot position and size are also dependent on the magnetic field.

To conclude, it is important to note that the synthetic data and inference done with it here rely on a simplified model. Care must be taken in the future analysis of real data, as results could be biased if a simplified or omitted piece of physics is significant.

\section{Conclusion}\label{sec:conclusion}
In this work, we first introduced updates related to \ac{AMP} physics to the existing \ac{X-PSI} infrastructure. These are the atmosphere model by \cite{Bobrikova2023} and an accretion disc. We also improved the computational efficiency of the code, which roughly cancelled the additional computational costs from these novel model additions.

In preparation for the analysis of real data, we tested \ac{PPM} of \acp{AMP} with two synthetic \ac{NICER} datasets and verified that the analysis pipeline is able to recover input parameters mostly accurately, but also found it depends on the \ac{AMP} \scenario. The two different \scenarios{} were inspired by \ac{J1808}, with special attention paid to the inference of mass and radius. In \largespot{}, we found tightly constrained parameters, comparable to results from a previous analysis with {\it RXTE} data by \cite{Salmi2018}. In \smallspot{} we found various degeneracies between parameters, including between disc and star properties. At least in part due to these degeneracies, some posteriors were widened compared to the previous \scenario. The recovered star mass was systematically lower than the true value, plausibly due to the bias on the inner disc radius caused by its prior. We concluded that special care must be taken in the modelling of the inner disc radius and determining the prior, to improve or safeguard the recovery of the star mass. In addition, joint observations that constrain either the star (polarization data for inclination and hotspot co-latitude) or accretion disc (e.g. broadened line reflection) would be valuable in this effort as well. Further research is needed to test how inference is affected upon the inclusion of a more sophisticated accretion disc model or additional constraints from joint observations. 

That two synthetic \scenarios{} provided different results indicates that the success of parameter inference will depend on the hotspot geometry, inclination and accretion disc geometry of the \ac{AMP}, meaning some \acp{AMP} are easier to constrain than others. The analyses done in this work provides a starting point for \ac{PPM} of real \ac{NICER} data of \acp{AMP}. 

\section*{Acknowledgements}

B.D. thanks Rudy Wijnands, Phil Uttley, Matteo Lucchini, Evert Rol and Martin Heemskerk for fruitful discussions and advice. B.D. thanks Duncan Galloway for interesting discussions and estimates for the distance of \ac{J1808}. B.D., T.S., A.L.W., Y.K., D.C., and S.V. acknowledge support from ERC Consolidator grant No. 865768 AEONS (PI: Watts). 
M.N. is a Fonds de Recherche du Quebec – Nature et Technologies (FRQNT) postdoctoral fellow.
J.P. thanks the Academy of Finland grant 333112 and the Ministry of Science and Higher Education grant 075-15-2024-647. 
A.B. was supported by the Finnish Cultural Foundation grant 00240328. 
This work used the Dutch national e-infrastructure with the support of the SURF Cooperative using grant no. EINF-5867. Part of the work was carried out on the HELIOS cluster including dedicated nodes funded via the above mentioned ERC CoG. We acknowledge extensive use of NASA’s Astrophysics Data System (ADS) Bibliographic Services and the ArXiv.

\section*{Software}
\texttt{X-PSI} \citep{Riley2023}, GNU Scientific Library (\texttt{GSL}; \citealt{Gough2009}), \texttt{HEASoft} \citep{HEASoft}, \texttt{MPI} for Python \citep{Dalcin2008}, \texttt{Multinest} \citep{Feroz2009}, \texttt{Pymultinest} \citep{PyMultiNest}, \texttt{nestcheck} \citep{Higson2018JOSS}, \texttt{GetDist} \citep{Lewis2019}, \texttt{Jupyter} \citep{2007CSE.....9c..21P, kluyver2016jupyter}, \texttt{astropy} \citep{astropy:2013, astropy:2018, astropy:2022}, \texttt{scipy} \citep{2020SciPy-NMeth, scipy_10909890}, \texttt{matplotlib} \citep{Hunter:2007}, \texttt{numpy} \citep{numpy}, \texttt{python} \citep{python}, \texttt{Cython} \citep{cython:2011} and \texttt{spyder} \citep{Spyder}.

\section*{Data availability}
A basic reproduction package for the analysis and all the figures are available at: \href{https://doi.org/10.5281/zenodo.13691859} {10.5281/zenodo.13691859}.

\bibliographystyle{mnras}
\bibliography{bibliography} 

\bsp	
\label{lastpage}
\end{document}